\documentclass[11pt]{article}
\usepackage[utf8]{inputenc}
\usepackage{times}
\usepackage{graphicx}
\usepackage{hyperref}

\usepackage{subcaption}
\usepackage{listings}
\usepackage{xcolor}
\usepackage{xspace}
\usepackage{lastpage}
\usepackage{booktabs}

\usepackage{paralist}
\usepackage{algpseudocode}
\usepackage{algorithm}
\let\oldReturn\Return
\renewcommand{\Return}{\State\oldReturn}

\usepackage[export]{adjustbox}

\newcommand{\namebig}[1]{Antidote SQL}
\newcommand{\name}[1]{AQL}

\newcommand{\antidote}[1]{Antidote}

\lstdefinestyle{basic}{
    basicstyle={\scriptsize\linespread{0.6}\ttfamily},
    numbers=left,
    numberstyle=\tiny\color{gray}\ttfamily,
    numbersep=5pt,
    backgroundcolor=\color{white},
    showspaces=false,
    showstringspaces=false,
    showtabs=false,
    frame=noe,
    rulecolor=\color{black},
    captionpos=b,
    keywordstyle=\color[rgb]{.4,.4,.6}\bf,
    commentstyle=\color{gray},
    stringstyle=\color[rgb]{.3,.5,.3}\bf,
    keywordstyle={[2]\color{black}\bf},
    tabsize=1,
}

\lstdefinelanguage{customJava}
{
    morekeywords={@forall,@Invariant,@Inv,@True,@False,@Increments,@Decrements,@Counter,forall,and},
    sensitive=false,
    morecomment=[l]{//},
    morestring=[b]",
    morestring=[b]',
}

\lstdefinelanguage{SQL}
{
    morekeywords={CREATE,TABLE,UPDATE_WINS,DELETE_WINS,PRIMARY,KEY,FOREIGN,REFERENCES,LWW,MULTI_VALUE,ADDITIVE,
    CHECK,ON,CASCADE,DELETE},
}

\usepackage[inline]{enumitem}

\begin{document}

%% remove copyright box
%\makeatletter
%\def\@copyrightspace{\relax}
%\makeatother

\title{\namebig{}: Relaxed When Possible, Strict When Necessary}
%\authorinfo{Valter Balegas \and Nuno Preguiça \and Sérgio Duarte \and Carla Ferreira }{NOVA-LINCS, Univ. Nova Lisboa}{}
%\authorinfo{Rodrigo Rodrigues}{INESC-ID \& IST, University of Lisbon}{}
%\author{Nuno Preguiça, Valter Balegas, Rodrigo Rodrigues, Annette Bieniusa, João Sousa, \\Pedro Lopes, Sérgio Duarte, Carla Ferreira}

\date{}

\author{Pedro Lopes$^1$, João Sousa$^4$\footnote{Work done while at NOVA LINCS, FCT, Universidade NOVA de Lisboa.}, Valter Balegas$^1$, Carla Ferreira$^1$, \\
Sérgio Duarte$^1$, Annette Bieniusa$^3$, Rodrigo Rodrigues$^2$, Nuno Preguiça$^1$\\
\\
$^1$ NOVA LINCS \& FCT, Universidade NOVA de Lisboa \\
$^2$ INESC-Id \& IST, University of Lisbon\\
$^3$ TU Kaiserlautern\\
$^4$ Feedzai}
\maketitle
\begin{abstract}

Geo-replication poses an inherent trade-off between low latency, high
availability and strong consistency.
While NoSQL databases favor low latency and high availability, relaxing
consistency, more recent cloud databases favor strong consistency and ease 
of programming, while still providing high scalability.
In this paper, we present \namebig{}, a database system that allows
application developers to relax SQL consistency when possible.
Unlike NoSQL databases, our approach enforces primary key, foreign key
and check SQL constraints even under relaxed consistency, which is
sufficient for guaranteeing the correctness of many applications. 
To this end, we defined concurrency semantics for SQL constraints under 
relaxed consistency and show how to implement such semantics efficiently.
For applications that require strict SQL consistency,
\namebig{} provides support for such semantics at the cost of requiring
coordination among replicas.

%Geo-replication poses an inherent trade-off between low latency, high
%availability and strong consistency.
%Some storage systems provide support for both weakly consistent and
%strongly consistent operations, leaving to the application developers
%the task of choosing the right consistency level for each operation.
%In this paper, we present \namebig{}, a database system that allows
%application developers to control the degree and outcome of concurrency in
%the database schema, including support for enforcing primary key, foreign key
%and check SQL constraints adopting optimistic and pessimistic approaches.
%We define sensible semantics for enforcing SQL constraints using an
%optimistic approach and show how to implement such semantics efficiently
%under weak consistency.
%For applications that require strict SQL semantics when accessing some data,
%\namebig{} provides support for such semantics at the cost of requiring
%coordination among replicas.

\end{abstract}

%========================================================================================
%========================================================================================
%========================================================================================
%========================================================================================

\section{Introduction}

% SQL was the standard for DB.. does not scale... eventual consistency 

SQL databases have been the \emph{de facto} standard for storing and managing data 
for many years. With the advent of cloud computing, and the need to scale applications 
to millions of users worldwide, new storage systems were designed that offered 
improved latency, availability and scalability over traditional SQL databases, giving rise to the NoSQL
movement~\cite{cassandra,dynamo}.

% hard to program... renewed interest in SQL - NewSQL

To provide such properties, these NoSQL systems exhibit some weaknesses:
\begin{inparaenum}[(i)]
  \item they only provide weak forms of consistency, which makes it difficult to ensure database integrity
  and application correctness;
  \item many of these systems only provide a key-value interface, which makes it difficult to model and query data efficiently.
\end{inparaenum}

These issues have led to a renewed interest in SQL, with the proposal of new designs that 
provide SQL semantics -- Spanner~\cite{spanner} , Aurora~\cite{aurora}, CosmosDB~\cite{cosmosdb} 
and VoltDB \cite{voltdb} are some recent examples of such databases. 
These systems offer practical high availability and scalability, but they are unable to ensure 
low latency at a global scale, as they rely on some form of consensus~\cite{paxos} to ensure 
consistency across sites.
For achieving low latency and high availability, it remains necessary to resort 
to weak consistency. 

% this paper: approach to relax SQL when possible

In this paper, we propose to allow programmers to relax SQL consistency 
when possible, while keeping stricter consistency when necessary.
Some systems \cite{Li2012RedBlue,dynamodb} provide an API 
with operations that run under weak or strong consistency, which could be used 
for this purpose. However, it has been 
shown that it is difficult to identify which operations need to execute under each 
consistency model, with several methodologies and tools being
proposed to help programmers in this process \cite{Li2012RedBlue,Li14Automating,indigo,cise,Roy14Writes}.

%The outcome is that applications are difficult to program because each database offers a different
%semantics, and in many cases it is unclear how to ensure consistency under that semantics.
%AQL offers a solution for this problem, allowing applications to take full advantage of the SQL
%standard and allowing programmers to specify which constraints can be relaxed in favor of
%better latency. It gives full control to programmers, as optimizations need to be defined
%explicitly in the database schema for each constraint that the programmer wants to relax.

We adopt a different approach: use the database schema to specify 
the degree of concurrency allowed.
With our concurrency-aware database schema, programmers identify which data items
can be modified concurrently and what should be the outcome of such concurrent
updates.
Additionally, they also specify which database constraints should be maintained 
and the degree of concurrency allowed while enforcing them.
The database system is then responsible for efficiently enforcing the defined
data model, minimizing the coordination used.
This approach gives full control to programmers, as they explicitly define when and
how SQL consistency can be relaxed.
In any case, our approach enforces database constraints, which is often 
sufficient for guaranteeing application correctness.
%which 
%concurrent accesses are possible, thus relaxing the SQL consistency.

%We adopt a different approach: use the database schema to specify 
%the degree of concurrency allowed and the constraints to maintain.
%With our concurrency-aware database schema, programmers identify which data items
%can be modified concurrently and what should be the outcome of such concurrent
%updates.
%Additionally, they also specify which database constraints should be maintained 
%and the degree of concurrency allowed while enforcing them.
%The database system is then responsible for efficiently enforcing the defined
%data model, minimizing the coordination used.
%This gives full control to programmers, as they should explicitly define when and
%how SQL consistency can be relaxed.
%In any case, our approach enforces database constraints, which is often 
%sufficient for guaranteeing application correctness.
%%which 
%%concurrent accesses are possible, thus relaxing the SQL consistency.

An important part of our work is the definition of sensible semantics
when handling concurrent updates.
For the outcome of concurrent updates to the same data items, we have built
on previous works \cite{Li14Automating,crdts-sss}, allowing programmers to
select the appropriate merge policy.
For supporting database constraints, including primary key, check and foreign
key constraints, we propose alternative semantics for dealing with 
concurrent updates.
While some semantics adopt an eventual consistency approach that poses no restriction
to concurrent updates by applying pre-defined merge policies, other semantics
restrict some concurrent updates.
Nonetheless, in the latter case, a high degree of concurrency is still possible.

% algorithms for implementing defined semantics

Implementing our approach, \namebig{}, in a geo-replicated setting, 
is challenging, as
data can be partitioned across multiple nodes in each data center.
First, enforcing referential integrity might involve relations between
data stored in different nodes, which could require
complex coordination among nodes for maintaining the database constraints.
We have devised a set of algorithms that avoid the need for coordination among
multiple nodes, thus leading to a simple and efficient solution.
Second, while adopting semantics that restrict concurrency, it is
important not to be over-restrictive.
Our proposed semantics and supporting algorithms achieve this goal.

% contributions

In summary, this paper makes the following contributions:

\begin{itemize}
    \item A database schema allowing to control when and how SQL
    consistency can be relaxed.
    \item The definition of sensible semantics for enforcing SQL constraints
    under weak consistency.
    \item A set of algorithms for enforcing the defined concurrency semantics.
\end{itemize}

\section{System overview}
\label{sec:system}

\namebig{} is designed for running in cloud infrastructures, composed
by multiple data centers, each one with multiple nodes.
Each data center fully replicates the database.
Inside each data center, data is sharded, with each shard being replicated
in a small number of nodes.

\namebig{} provides a SQL-like interface, \name{}, to applications.
Applications define the database schema using the \name{} data definition language (DDL).
\name{} DDL extends SQL DDL by allowing programmers to specify
the concurrency semantics for the database.
This concurrency semantics includes specifying what concurrency is allowed when
accessing the database and what should be the outcome of concurrent updates.

Applications access the database by issuing transactions that include a sequence
of standard SQL statements, including the \texttt{select} statement for
querying the database and \texttt{insert}, \texttt{update} and \texttt{delete}
statements for updating the database.

\name{} transactions run under parallel snapshot isolation (PSI)
semantics \cite{walter} extended with integrity constraints.
PSI is a an extension of snapshot isolation (SI) for geo-replicated
settings.
As SI, PSI precludes write-write conflicts between concurrent transactions,
unless they are writes to mergeable data types.
However, unlike SI, PSI allows different sites to order transactions differently,
if the order preserves causal ordering: if a transaction T2 reads from
T1, then T1 must be ordered before T2 at every data center.
Under PSI, all operations of a transaction running in a given site,
read the most recent committed version at that site as of the time
of transaction begin.

We extend PSI to enforce integrity constraints. Under this model, at every site,
all snapshots preserve the integrity constraints defined in the
database schema, including primary key, check and foreign key constraints.
As discussed later, integrity constraints can be enforced using both
optimistic and pessimistic approaches, with the former being a highly
available solution that solves conflicts according to the user defined policy
(see Section~\ref{sec:integirty-constraints-semantics}).
%\annette{Maybe add: Constraint violations do not lead to an abort / rollback,
%but are resolved according to a user defined policy (see sec 3.2.).}

%========================================================================================
%========================================================================================
%========================================================================================
%========================================================================================
\section{Concurrency Semantics}

\namebig{} allows programmers to control the allowed concurrency among
transactions through the database schema.
When concurrency is allowed, an important aspect is the concurrency semantics,
which defines the outcome in the presence of concurrent updates.
This section discusses the supported concurrency semantics.

\subsection{Database Model}

\namebig{} supports a relational data model, where data is stored in tables
with a given schema. We now present the options for controlling
concurrency associated with each table.

\emph{Semantics for update-delete:}
When creating a table, programmers can specify
whether it will be possible to concurrently update and delete a table row.
\name{} provides three possible semantics (Figure~\ref{fig:aql:create:table}): \emph{update-wins},
\emph{delete-wins} and \emph{no concurrency} (if no modifier is specified).
In the \emph{update-wins} semantics, when concurrent transactions execute a delete
and an update operation over the same row, the effects of the delete over that row
are ignored.
In the \emph{delete-wins} semantics, the effect of the delete will prevail and the row
is deleted.
In the \emph{no concurrency} semantics, concurrent transactions cannot execute
a delete and an update operation over the same row.

The first two semantics lead to a \emph{lost update}, as one
of the operations will be ignored. However, the final state of the database
depends only on the type of operations concurrently executed, and not on
an arbitrary order among updates established at runtime, as it is the
case for example in last-writer-wins solutions \cite{cops}.
%\annette{Hmm, the result depends (non-deterministically) on the concurrency pattern.}

\begin{figure}
\begin{lstlisting}[language=SQL,frame=single,basicstyle=\small\ttfamily]
CREATE [UPDATE_WINS|DELETE_WINS] TABLE table_name(
  column1 datatype [constraint],
  column2 datatype [constraint],
  ...
  column_n datatype [constraint]
)
\end{lstlisting}
\caption{\name{} create table statement.}
 \label{fig:aql:create:table}
\end{figure}

\emph{Semantics for update-update:}
Programmers can specify which updates can be made concurrently to the same row
when defining the table schema.
To this end, \name{} provides the following modifiers for the columns (Figure~\ref{fig:aql:datatype:modifiers}):
\emph{last-writer-wins}, \emph{multi-value} and \emph{additive}.

\begin{figure}
\begin{lstlisting}[language=SQL,frame=single,basicstyle=\small\ttfamily]
generic_modifier ::= LWW | MULTI_VALUE
numeric_modifier ::= generic_modifier | ADDITIVE
\end{lstlisting}
\caption{Modifiers for \name{} data types.}
 \label{fig:aql:datatype:modifiers}
\end{figure}

In the \emph{last-writer-wins} semantics, when concurrent updates modify the
same row, the value of the last update (as ordered according to the wall clock)
will prevail.
In the \emph{multi-value} semantics, when concurrent updates modify the
same row, the database will store both values.
This option should be used carefully, as it will affect the result returned
by select operations, with multiple values being returned (instead of a single one).
Finally, the \emph{additive} semantics, for being used with numeric data types,
allows the final state to merge all updates to the numeric value.
Thus, given two concurrent update operations that add \emph{$k_1$} and \emph{$k_2$}
to a column, the final database state will have the initial value of the
column incremented by \emph{$k_1 + k_2$}.

If no modifier is used for a given column, the system will not
allow concurrent updates that modify this column in the same row.
Updates that modify this column in different rows are allowed.

The semantics of update-update also control whether it is possible to
concurrently insert a row with the same primary key.
If all columns (besides the primary key) have a concurrency modifier,
concurrent inserts are allowed, with the final state
being determined by using the the defined semantics for each column.

\subsection{Integrity Constraints}
\label{sec:integirty-constraints-semantics}

In the previous section, we presented the options for controlling concurrency
among multiple clients by restricting concurrent updates to the same data items
or adopting appropriate merge policies.
We now present the semantics for controlling concurrent accesses that may lead to a
database constraint violation (Figure~\ref{fig:aql:constraints} presents
the syntax for specifying constraints).
\begin{figure}
\begin{lstlisting}[language=SQL,frame=single,basicstyle=\small\ttfamily]
constraint ::=
    PRIMARY KEY |
    CHECK (condition) |
    FOREIGN KEY [UPDATE_WINS|DELETE_WINS]
      REFERENCES table(column) [ON DELETE CASCADE]
\end{lstlisting}
\caption{Integrity constraints supported by \name{}.}
 \label{fig:aql:constraints}
\end{figure}

\emph{Primary key constraint:}
The primary key constraint is used to guarantee that the value of the primary key column
is unique for each row in the table.
We support two alternative approaches to guarantee this constraint.

First, if some column of the table (other than the primary key) includes the no
concurrency semantics, no concurrent inserts will be allowed.

Second, if all columns (besides the primary key) include a concurrency semantics,
\name{} will allow multiple insert operations to be executed concurrently.
The final value of each column is determined according to its concurrency semantics.

Both approaches guarantee that a single row with a given primary key exists,
with the former restricting concurrency.
One practical aspect that is important for primary keys is how applications concurrently
generate different primary keys. To this end, \name{} provides two functions, one returning
a unique identifier and the other a sequential unique identifier (encoded as a number).

%\emph{Unique constraints:}
%
%

\emph{Check constraint:}
The check constraint allows to specify that the value of a column respects some
given condition. For example, the check constraint can be used to guarantee that
the stock of some product is not negative.

\name{} allows programmers to specify check constraints for any column.
For numeric \emph{additive} columns, \name{} allows the value of the column to be updated
concurrently, when it is possible to guarantee that the updates will not
make the condition false.
As detailed later, to support this constraint, our prototype relies on
escrow techniques \cite{ONeil1986Escrow}.

A transaction running at a site aborts if an update that modifies the value of a column
that has a check constraint may lead to an invariant violation.
The database will return to the application information that allows the programmer to
know if the transaction might commit if retried.

\emph{Foreign key constraint:}
The foreign key constraint allows to relate entries from different tables, by making the values
of a column in one table uniquely identify rows in some other table.

\begin{figure}
\begin{center}
\includegraphics[width=\columnwidth]{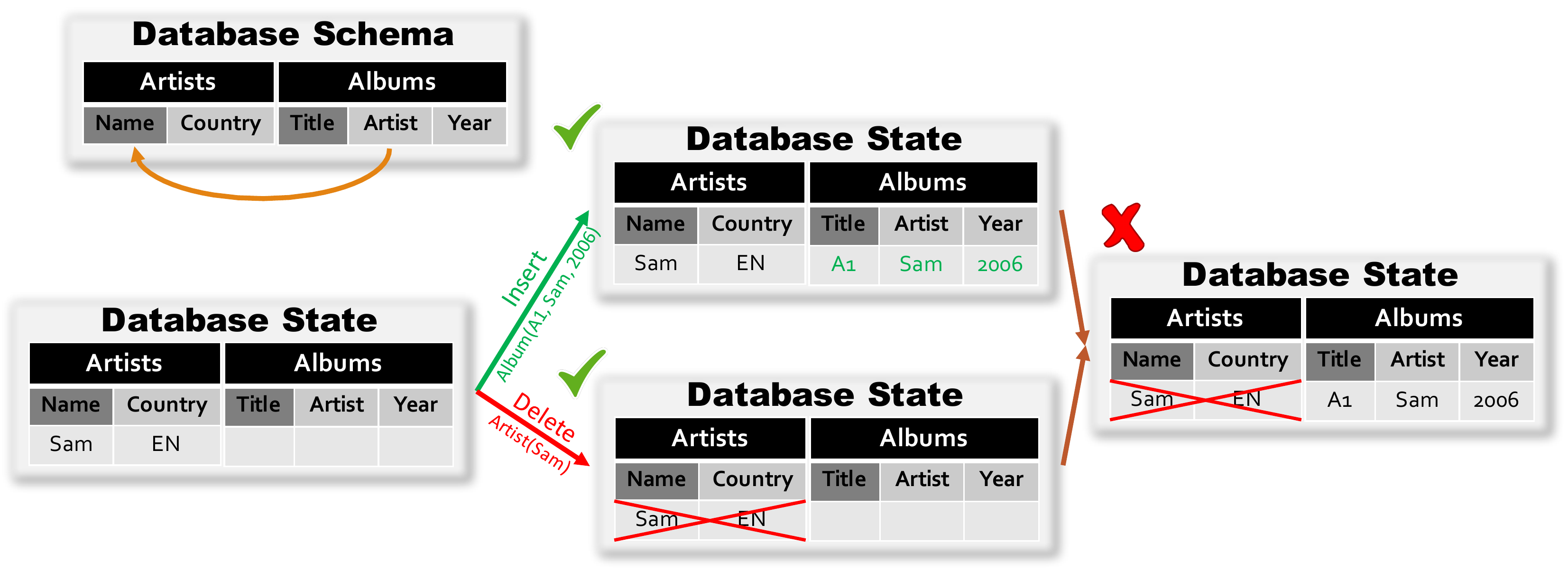}
 \caption{Example of foreign key constraint violation.}
 \label{fig:aql:fk:error:simple}
\end{center}
\end{figure}

Foreign key constraints are particularly challenging in our system model, as a 
constraint violation can result from concurrent
updates in different tables.
Consider the example of Figure~\ref{fig:aql:fk:error:simple}.
In this example, the database contains two tables,
Artists and Albums, where Albums has a foreign key in column Artist that references
the Artists table.
In the example, starting in a database state where only
artist \emph{Sam} exists, two transactions concurrently delete the artist
and add an album for the artist.
When combining the effects of the two transactions, we would reach a state with an album referring
to a deleted artist, leading to a violation of the foreign key constraint.

\name{} supports the following concurrency semantics for handling updates that
affect a foreign key
constraint: \emph{update-wins} semantics, \emph{delete-wins} semantics and
\emph{no concurrency}.

\begin{figure}[t]
\centering
\begin{subfigure}{\columnwidth}
  \centering
\includegraphics[width=\textwidth]{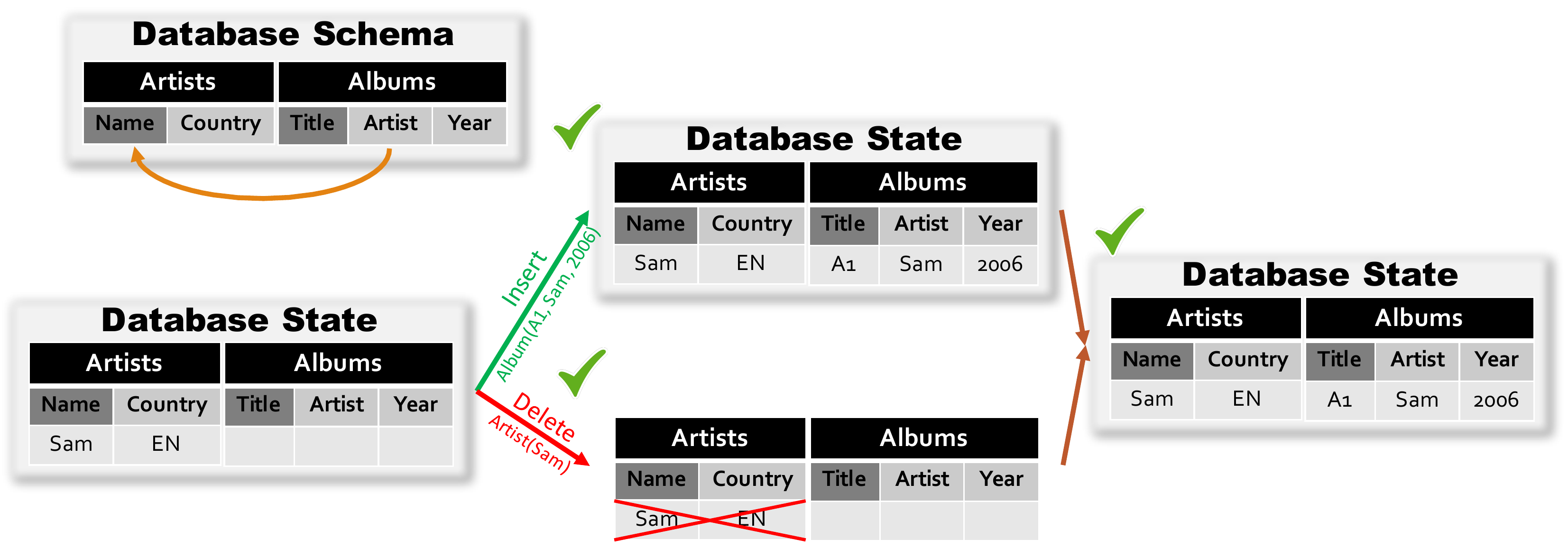}
 \caption{\emph{Update-wins} semantics.}
 \label{fig:aql:fk:error:simple:revive}
\end{subfigure}
\\
\begin{subfigure}{\columnwidth}
  \centering
\includegraphics[width=\textwidth]{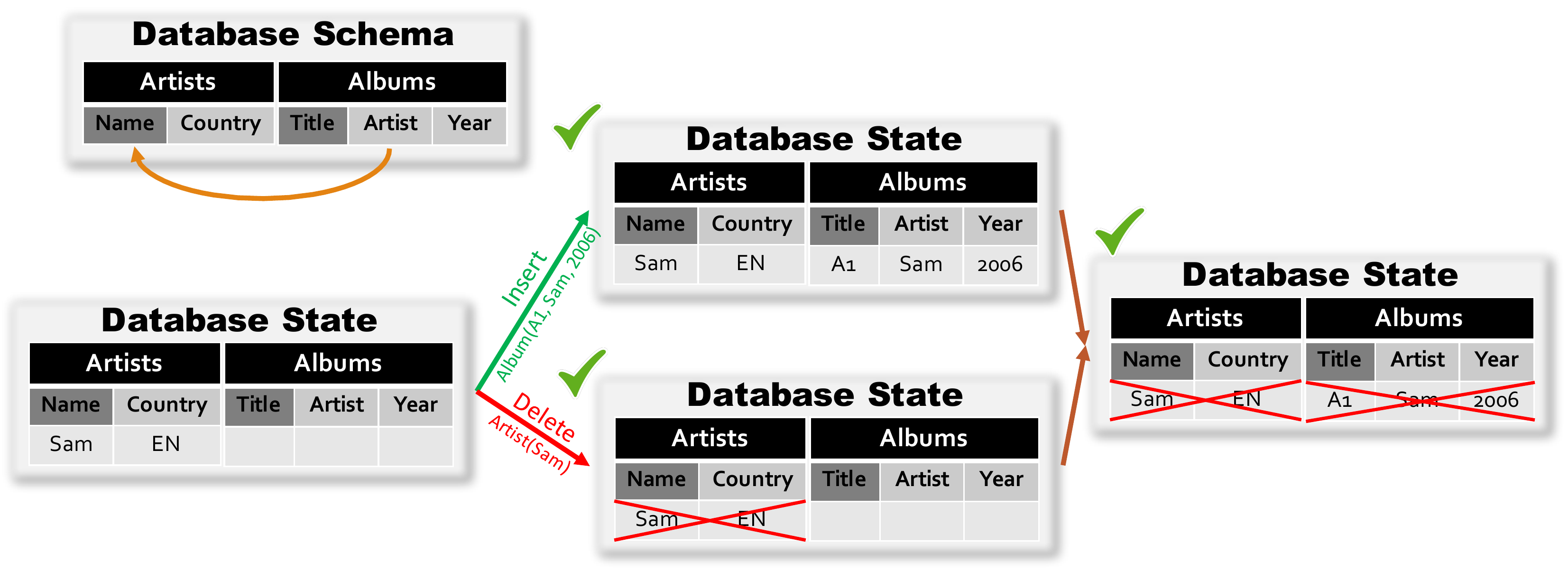}
 \caption{\emph{Delete-wins} semantics.}
 \label{fig:aql:fk:error:simple:norevive}
\end{subfigure}
\caption{Semantics for solving foreign key constraint violations.}
\label{fig:aql:fk:error:simplesolution}
\end{figure}

In the \emph{update-wins} semantics, when concurrently deleting row $r$
and inserting a row that references row $r$, the delete has no effect
in the final database state -- Figure~\ref{fig:aql:fk:error:simple:revive}
shows the effect of \emph{update-wins} in the previous example.
Conversely, in the \emph{delete-wins} semantics, it is the insert
operations that will have no effect in the final database state --
Figure~\ref{fig:aql:fk:error:simple:norevive}
shows the effect of \emph{delete-wins} in the previous example.

In the \emph{no concurrency} semantics, the system will not allow the concurrent
deletion of a row $r$ and the insertion of some row that references $r$.
We note that in this case, it is still possible to have multiple concurrent
inserts that reference the same row.

\begin{figure}
\begin{center}
\includegraphics[width=\columnwidth]{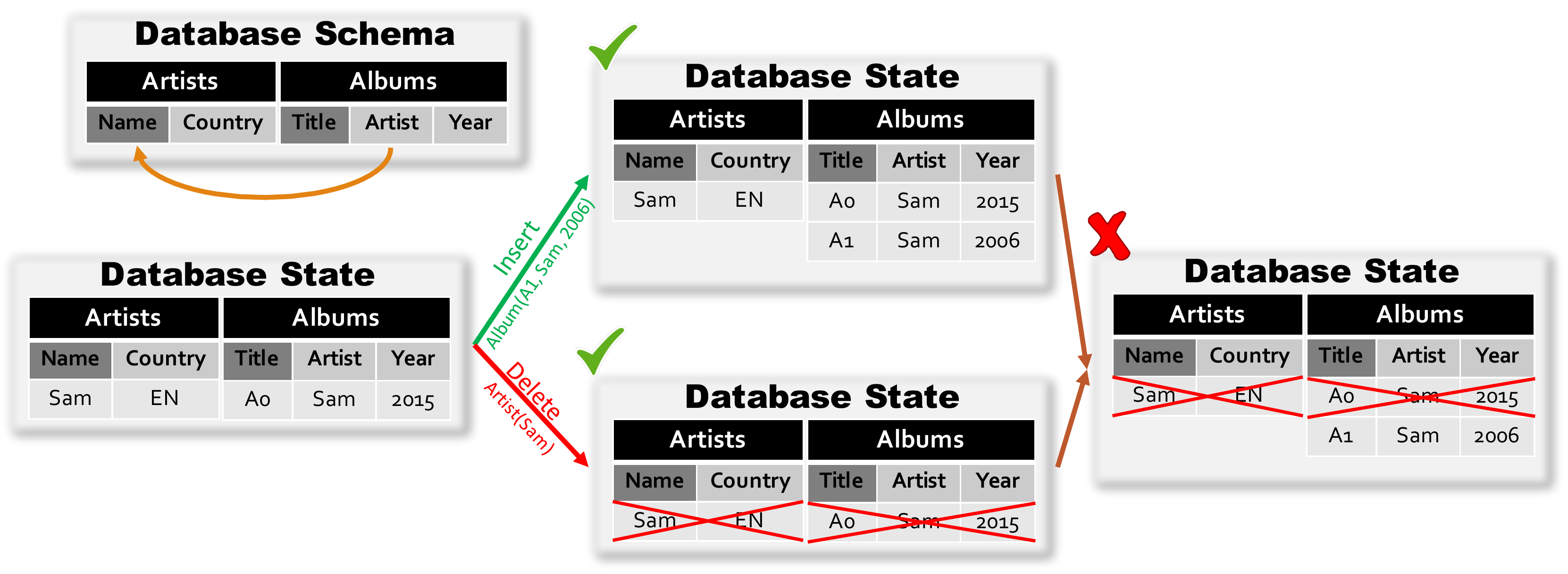}
 \caption{Example of foreign key with cascading constraint violation.}
 \label{fig:aql:fk:error}
\end{center}
\end{figure}

We now discuss the case when a foreign key is defined with the \emph{on cascade
delete} behavior.
Consider the example of Figure~\ref{fig:aql:fk:error}, that starts with
a database state including artist \emph{Sam} with an album \emph{A0}.
A transaction adds album \emph{A1} for \emph{Sam}, while a concurrent transaction
deletes artist \emph{Sam}. The cascading effect leads
to the deletion of album \emph{A0} (that was the only known album in the site where
the delete was executed).
Combining the effects of both transactions leads to a foreign key constraint
violation with album \emph{A1} referring to the deleted artist \emph{Sam}.

\begin{figure}
\centering
\begin{subfigure}{\columnwidth}
  \centering
\includegraphics[width=\textwidth]{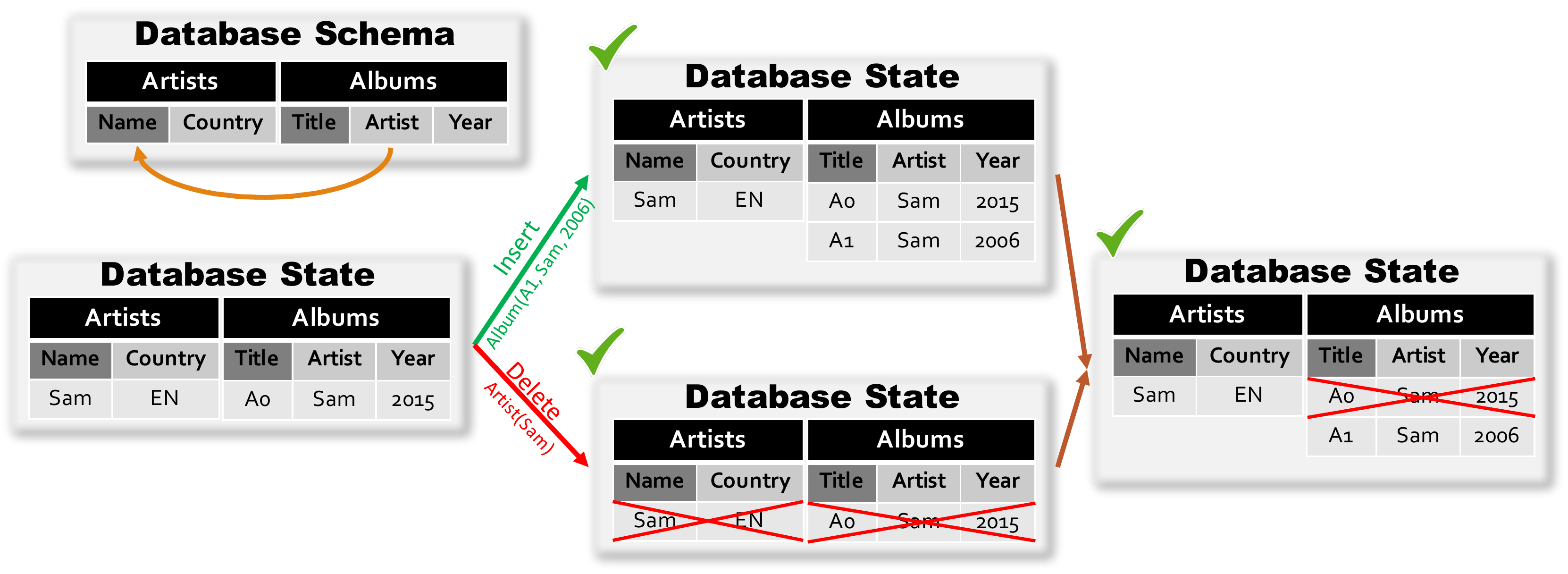}
 \caption{\emph{Update-wins} semantics.}
 \label{fig:aql:fk:error:revive}
\end{subfigure}
\\
\begin{subfigure}{\columnwidth}
  \centering
\includegraphics[width=\textwidth]{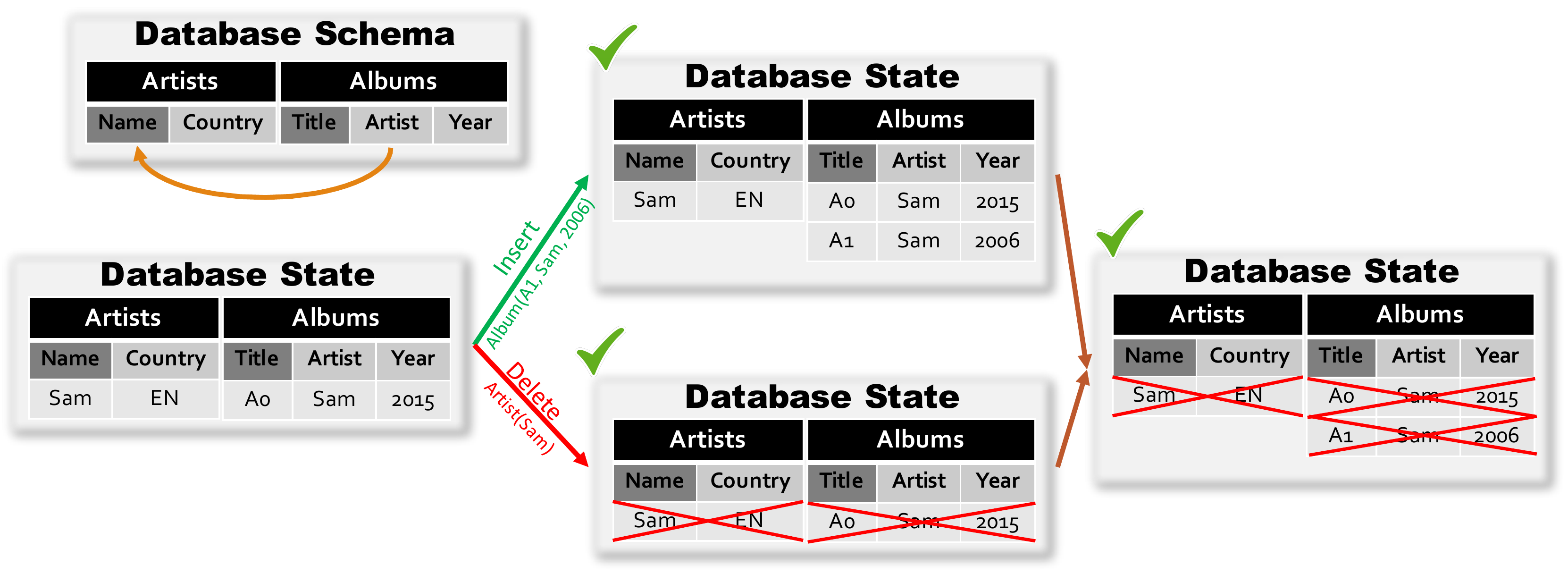}
 \caption{\emph{Delete-wins} semantics.}
 \label{fig:aql:fk:error:norevive}
\end{subfigure}
\caption{Semantics for solving foreign key constraint violations.}
\label{fig:aql:fk:errorsolution}
\end{figure}

With cascading, the \emph{delete-wins} semantics has exactly the same
behavior as before, making the concurrent insert to have no effect -- as
shown in Figure~\ref{fig:aql:fk:error:norevive}, the final database state
does not include album \emph{A1}.
%The rational is that the cascading effect also applies to the concurrent operations
%that affect the Albums table.

For the \emph{update-wins} semantics, different alternatives could be
considered.
First, the delete operation could have no effect -- in this case, the final database
state would include both albums \emph{A0} and \emph{A1}.
Second, for the delete operation, only the effects that would lead to a
foreign key violation would be ignored -- in this case, the final
database state would include only album \emph{A1}.
We chose the latter option, as it is the one where less effects are ignored
-- Figure~\ref{fig:aql:fk:error:revive} exemplifies this case.
The general rule adopted in \name{} is the following: when the effects of an operation
are ignored due to a concurrent operation, we try to minimize the effects
ignored.

\subsection{Discussion}

\name{} allows programmers to have full control of when and how to 
relax SQL semantics by specifying the degree and outcome of concurrency 
allowed in the database schema.
For data that is critical to application correctness, the programmer 
can select to forbid concurrent accesses, thus keeping strict SQL consistency,
or to allow concurrent accesses given that database constraints
are maintained using appropriate semantics.

For example, consider a database for an
on-line shop. For guaranteeing that some product is not oversold, the
programmer can use a check constraint that achieves this goal while allowing
concurrent updates to commit while there is plenty of stock available.
%In this case, when a transaction executes, the transaction has no access to
%the exact value of the available stock. 
%If this is important, the programmer can prevent concurrent accesses to the
%stock, which would restrict concurrency.

As in any other database, one can expect that a large number of foreign key 
constraints exist. 
Consider the following foreign keys for a shopping cart: 
the shopping cart refers to the client that owns it; a shopping cart entry refers
to a shopping cart and to a product.
The programmer can use the \name{} \emph{no concurrency} semantics to prevent concurrent updates 
that would break the foreign key constraints and ultimately the application correctness.
%In this case, it would be impossible to delete a shopping cart and adding new entries concurrently, 
%but it would be possible to add entries concurrently.
Alternatively, she could use the \emph{delete-wins} semantics to guarantee that,
when a shopping cart is deleted, all information associated with the shopping cart is 
also deleted despite any concurrent updates. This would typically lead to the
expected result for the application.

\section{Algorithms and Prototype}

The concurrency semantics presented in the previous section allow programmers
to control the degree of concurrency allowed in their applications and to reason
about the behavior of their applications when concurrency is allowed.
In this section, we briefly present the algorithms we have developed for efficiently
implementing the proposed concurrency semantics in a geo-replicated setting, where
data is partitioned across multiple nodes in each data center.

\namebig{} is the SQL interface for AntidoteDB\footnote{\url{http://antidotedb.org}}, a geo-replicated transactional
key-value store with CRDT objects \cite{crdts-sss} and highly-available transactions via PSI (see Section \ref{sec:system}).
% AntidoteDB provides highly-available transactions with PSI for mergeable objects \cite{antidote}.
% Under this consistency model, a transaction runs against the most recent stable snapshot
% in the local data center. A transaction can always commit, with concurrent updates to
% the same objects being merged according to the semantics of the CRDT data type.
% \annette{Duplication with sec 2 on PSI}
For mapping the relational data to AntidoteDB's interface and supporting SQL
operations efficiently, our prototype uses techniques that have been employed in
other SQL interfaces for key-value stores.
Each row of a table is mapped to a key/value pair, where the key is
built from the table name and primary key, and the value stores the contents of
the row.
For supporting queries efficiently, our prototype maintains a primary key index
and secondary indexes (if the programmer creates such indexes).
We now focus on how to support the \name{} concurrency semantics efficiently.
Due to space limitations, we omit here the aspects related with
the interaction between index maintenance and concurrent updates to indexed values,
and with garbage-collection, that is performed asynchronously.

\emph{Multi-level locks:}
For supporting the \emph{no concurrency} semantics, our prototype resorts to a distributed
implementation of a multi-level lock (MLL) \cite{indigo}, with two modes: shared and exclusive.
Each lock is controlled in two levels.
First, the lock can be owned in exclusive mode by a single data center
or in shared mode by any set of data centers.
Second, an exclusive lock owned by a data center can be acquired by a single transaction
running in that data center. A shared lock can be acquired by multiple transactions.

\subsection{Database Model}

\emph{Update-delete semantics:}
The \emph{no concurrency} semantics is implemented by requiring a transaction to acquire:
\begin{inparaenum}[(i)]
\item in shared mode, the locks for the primary keys of the rows modified by an
update operation; and
\item in exclusive mode, the locks for the primary keys of the rows deleted by
a delete operation.
\end{inparaenum}

For supporting the \emph{update-wins} and \emph{delete-wins} semantics, we
use an hidden column (visibility column) in each row to control whether the row
has been deleted or not.
When a delete operation is executed, the column is assigned the value \texttt{D}.
When the row is updated (or inserted), the column is assigned the value \texttt{I}.
This column is implemented using a multi-value register CRDT, that stores all
values assigned concurrently to the register.
Thus, when an update operation executes concurrently with a delete operation for the same
row, the final value of the visibility column will include both \texttt{D} and \texttt{I}.
For a table with the \emph{update-wins} semantics, a row is considered as deleted if and
only if \emph{the only value} of the visibility column is \texttt{D}.
For a table with the \emph{delete-wins} semantics, a row is considered as deleted if and
only if \emph{one of the values} of the visibility column is \texttt{D}.

\begin{figure}
\begin{center}
\includegraphics[width=\columnwidth]{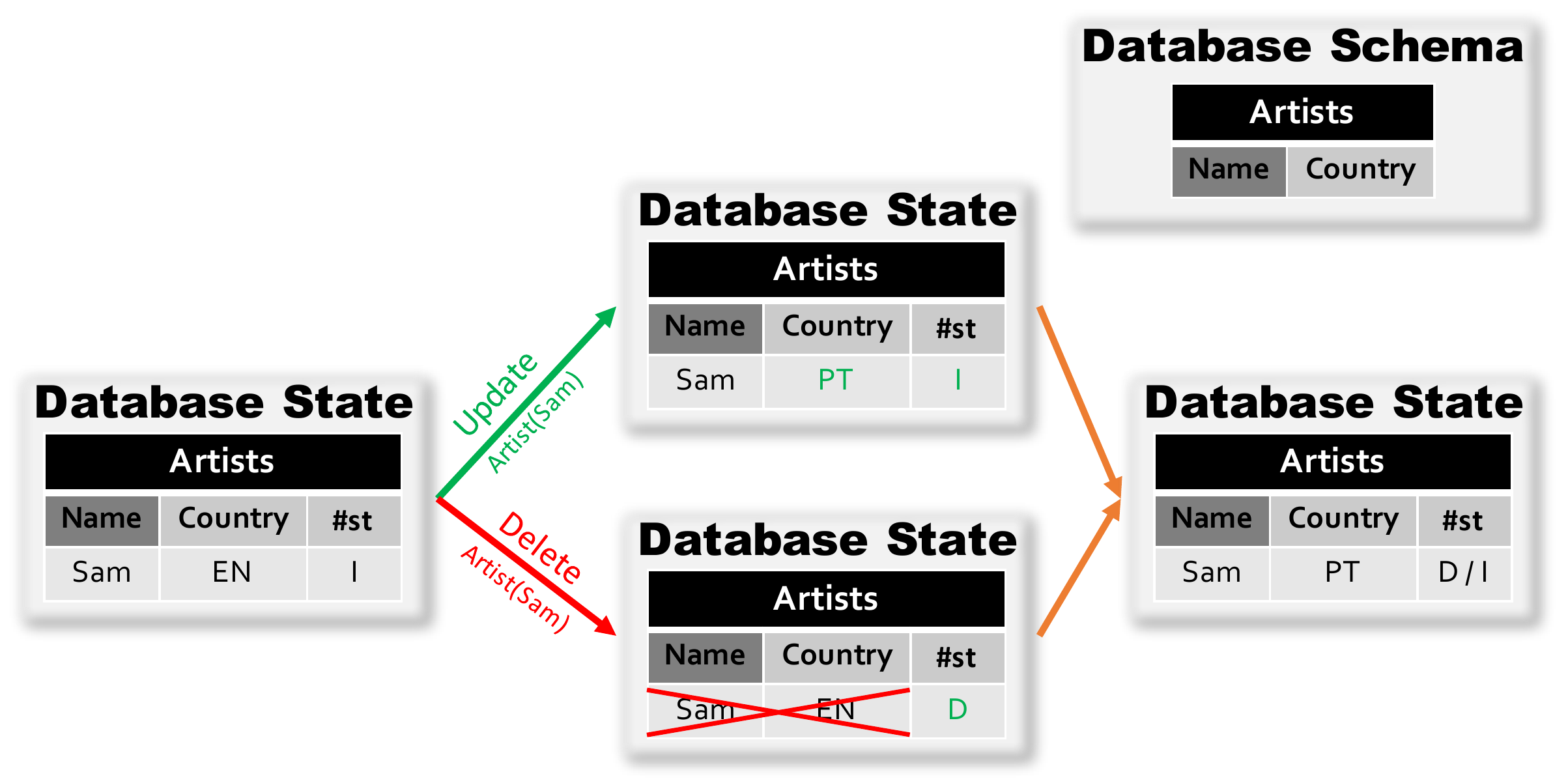}
 \caption{Example of the evolution of the visibility column for concurrent update and delete operations.}
 \label{fig:aql:delete}
\end{center}
\end{figure}

\emph{Update-update semantics:}
For supporting the merge policies associated with each column, we build on
the CRDTs supported by AntidoteDB.
Thus, the \emph{last-write-wins} semantics is implemented by storing the
value of the column in a last-writer-wins register CRDT.
The \emph{multi-value} semantics is implemented by storing the
value of the column in a multi-value register CRDT.
The \emph{additive} semantics is implemented by storing the value of the
column in a counter CRDT.

For supporting the \emph{no concurrency} semantics for a column, we associate a lock
with the primary key and column name. An update operation that modifies
the column must acquire the lock in exclusive mode before proceeding.

\subsection{Integrity Constraints}

\emph{Primary key constraint}: Primary key constraints can be enforced with
two different approaches, as explained in Section~\ref{sec:integirty-constraints-semantics}.
In the case that some column uses the \emph{no concurrency} semantics, we use MLLs 
to enforce mutual exclusion. 
For generating sequential unique identifiers we also use a MLL.
For (non-sequential) unique identifier, the identifiers are generated with a prefix
per site to avoid identifier collisions.

In the second case, where every column (except the primary key) uses some concurrency
semantics, no mechanism for preventing concurrent inserts is necessary, as each column in the row
can be merged with the specified concurrency semantics.

\emph{Check constraints}: To support check constraints, for all columns other than numeric
\emph{additive} columns, it suffices to check that the column value conforms to the specified
condition when a row is inserted or the column is updated.

For columns with the \emph{additive} semantics, \namebig{} relies on the
bounded counter CRDT \cite{bcounters} available in AntidoteDB.
The bounded counter CRDT implements the Escrow model~\cite{ONeil1986Escrow}:
permissions are granted to each holder of the counter (a replica) to execute operations
without coordination as long as the local delta on the value of the counter does not exceed
some threshold (and the sum of all thresholds still meets the defined condition).
This ensures that after propagating the deltas executed in each replica,
the value of the counter always meets the defined constraint.
If some replica needs to exceed its current threshold, it can negotiate with another replica
to change its threshold.

%For non-numerical types we use register CRDTs that
%merge concurrent updates based on their recency, or by establishing a total order of values.
%For numerical types, we support different the bound types ($<$, $\leq$, $>$, $\geq$).
%
%For the \emph{no concurrency} semantics, we use a MLL as before.
%For the \emph{additive} semantics, AQL builds on bounded counter CRDTs\cite{bcounters}.
%The bounded counter implements the Escrow model~\cite{ONeil1986Escrow}, which consists in
%granting permissions to each holder of the counter (a replica) to execute operations
%without coordination as long as the local delta on the value of the counter does not exceed
%some threshold. This ensures that after propagating the deltas executed in each replica,
%the value of the counter always meets the defined constraint.
%If some replica needs to exceed its current threshold, it can negotiate with another replica
%to change its threshold.

%\begin{table}[h]
%\label{tab:constraint:check}
%\caption{Mapping between bound types and their correspondent $btype$ and $isEq$ values.}
%\centering
%\begin{tabular}{c|c|c}
%\hline
%Bound type & \textit{btype} & \textit{isEq} \\
%\hline
%Lesser     & -1             & 1    \\
%LesserEq   & -1             & 0    \\
%Greater    & 1              & 1    \\
%GreaterEq  & 1              & 0    \\
%\hline
%\end{tabular}
%\end{table}

%\begin{figure}
%\begin{center}
%\includegraphics[width=0.5\textwidth]{images/aql-fk-flags.pdf}
% \caption{Evolution of visibility flags when inserting a new row with a foreign key (without cascading).}
% \label{fig:aql:fk:flags}
%\end{center}
%\end{figure}

\emph{Foreign key constraint}:
When designing the algorithms for enforcing the different semantics
supported for the foreign key constraint, an important aspect to consider is that,
as shown in Figure~\ref{fig:aql:fk:error:simple}, a conflict may occur due to
updates performed in different tables. Thus, it is not possible to detect
a conflict simply by checking the occurrence of concurrent updates to the same
data item.

For the \emph{no concurrency} semantics, we use MLLs to control concurrent accesses
that could break the foreign key constraint, requiring a transaction to acquire: 
\begin{inparaenum}[(i)]
\item an
exclusive lock for deleting a row in the parent table; and
\item a shared lock on the parent table 
for inserting a row in the child table.
\end{inparaenum}
Thus, in our running example, a delete in table Artists will require an exclusive lock
for the primary keys of the deleted rows; an insert (or update) in the table Albums
requires a shared lock for the primary key of the referenced row.
We note that this approach allows insertions to execute concurrently -- in many applications,
this will be the general case, thus enabling transactions to proceed concurrently in multiple
data centers.

\begin{figure}
\begin{center}
\includegraphics[width=\columnwidth]{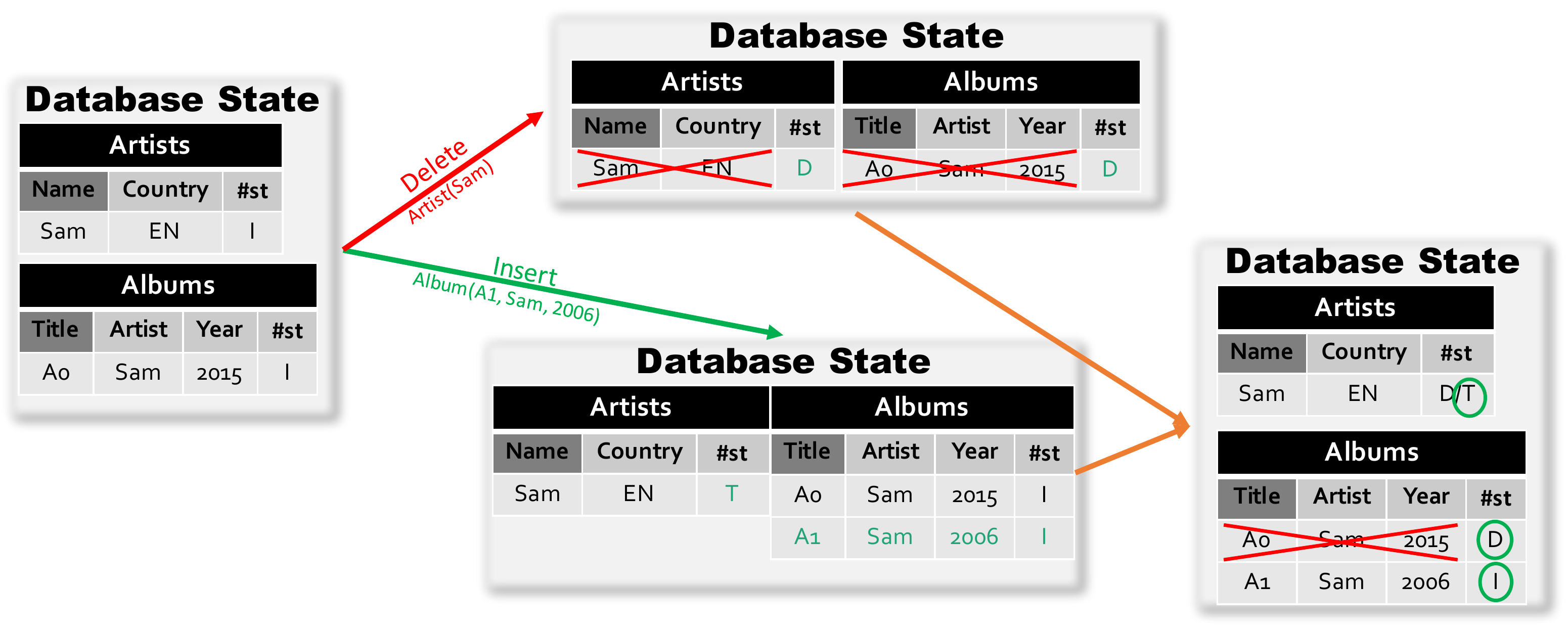}
 \caption{Evolution of visibility flags for \emph{update-wins} foreign key constraint (example of Figure~\ref{fig:aql:fk:error:revive}).}
 \label{fig:aql:fk:error:revive:flags}
\end{center}
\end{figure}

For implementing the \emph{update-wins} semantics, we resort to the visibility flags
associated with each row and extend the effects of insert operations.
Figure~\ref{fig:aql:fk:error:revive:flags} exemplifies the algorithm implemented, using the
example previously presented in Figure~\ref{fig:aql:fk:error:revive}.
A delete of a row that has no child will succeed, with the row marked as deleted by
setting its visibility flag to \texttt{D}.
If the row is referenced by other rows, the delete will only succeed if the foreign key
constraint was declared as delete on cascade. In this case, both the parent and child
rows are marked as deleted.
This can be seen in Figure~\ref{fig:aql:fk:error:revive:flags}, with the deletion of artist
\emph{Sam} (and cascading delete of album \emph{A0}).

When inserting a row that references another row, we mark the parent row as touched, by
setting its visibility flag to \texttt{T} -- in our example, the insertion of Album
\emph{A1} sets the visibility flag of artist \emph{Sam} to \texttt{T}.
By making the visibility flag \texttt{T} stronger than \texttt{D}, we can make
sure that in this case the parent row will not be deleted.
In our example, when merging the concurrent updates, the visibility flags
associated with artist \emph{Sam} include both \texttt{T} and \texttt{D}.
For \emph{update-wins} foreign key semantics, where \texttt{T} is stronger than
\texttt{D}, this means that the row is visible (a row is visible unless its visibility
flag is only \texttt{D}).
For the albums, album \emph{A0} remains deleted and album \emph{A1} is visible, as
defined in our \emph{update-wins} semantics.

\begin{figure}
\begin{center}
\includegraphics[width=\columnwidth]{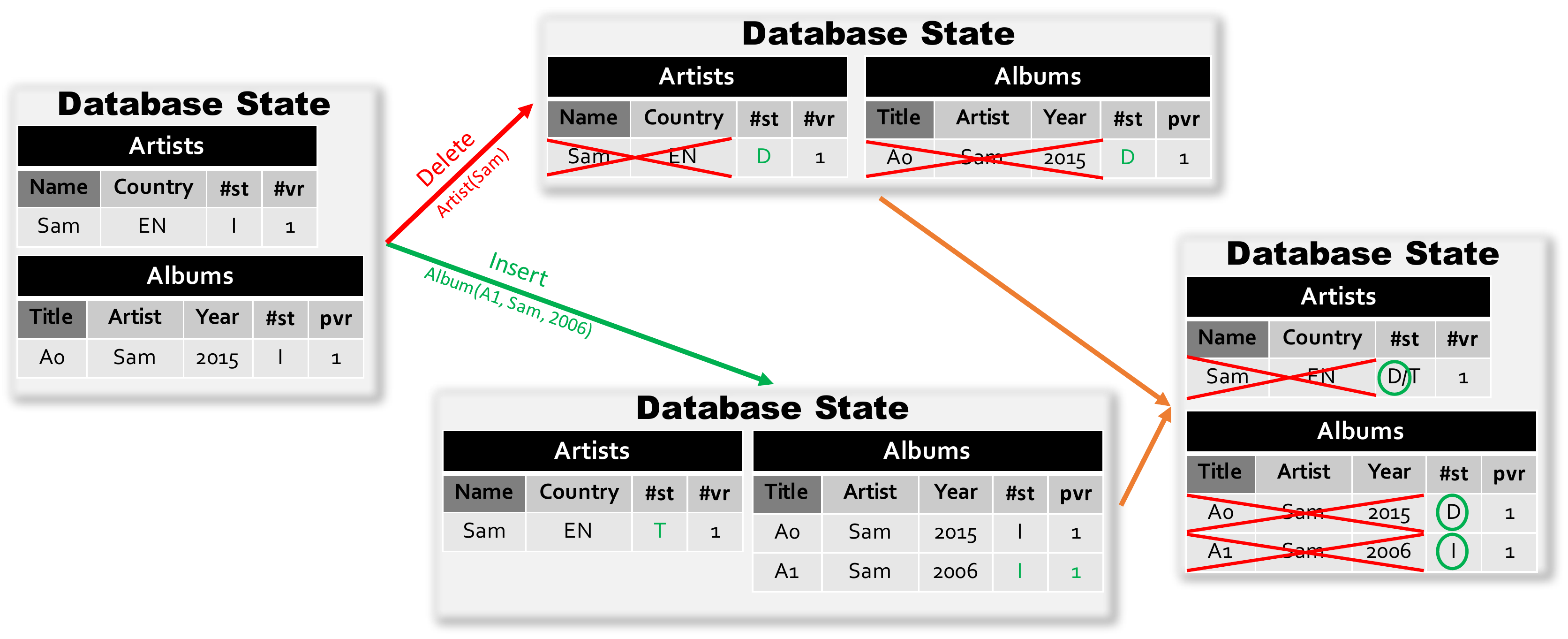}
 \caption{Evolution of metadata for \emph{delete-wins} foreign key constraint (example of Figure~\ref{fig:aql:fk:error:norevive}).}
 \label{fig:aql:fk:error:norevive:flags}
\end{center}
\end{figure}

Implementing the \emph{delete-wins} semantics is more complex.
While in the \emph{update-wins} semantics it was possible to enforce the
undo of the effects of the delete easily by forcing a conflict between the
delete and the touch in the parent row, this is not possible in the
\emph{delete-wins} semantics. In this case, on a concurrent insertion of a child
and the deletion of the parent row, the inserted child is not known at the time
of the deletion.

To achieve the intended semantics, besides using the visibility flags, we
extend read operations to check if the parent row has been deleted or not
(often, the value was already read in the transaction and no additional read
needs to be executed).
Figure~\ref{fig:aql:fk:error:norevive:flags} shows our running example.
In this case, when reading from table \emph{Albums}, if the row is visible,
it is necessary to check if the parent Artist is also visible.

We note that in this case, besides the visibility flags, we also maintain a
version identifier for the parent rows. This is necessary to guarantee that
if the element in the parent row is reinserted (after being deleted), a
deleted child row is not visible again -- in our example, album \emph{A1}
is only visible if the parent version 1 (\emph{pvr}) of artist \emph{Sam} is visible.

\section{Related work}
\label{sec:relatedwork}

Geo-replication has become a key feature in cloud storage systems, with data being
replicated in multiple data centers spread around the world.
The goal of geo-replication is to provide high availability and low latency,
by allowing clients to access any nearby replica.
To achieve these properties, a number of systems \cite{dynamo,cops} adopt a weak
consistency model, where an update can execute in any replica, being propagated
asynchronously to other replicas.

Writing correct applications under weak consistency can be complex.
To address this problem, several geo-replicated storage systems \cite{spanner,aurora,cosmosdb} 
adopt a strong consistency approach.
While several optimization techniques have been proposed for improving throughput \cite{spanner}
and latency \cite{dpaxos}, executing operations involves inter-data-center coordination,
with impact on latency and availability.

Our work is closer to systems \cite{Li2012RedBlue,dynamodb} 
that provide support for both weak and strong consistency.
For helping programmers decide which operation should execute under each consistency model, 
several tools have been proposed \cite{Li2012RedBlue,Li14Automating,indigo,cise,Roy14Writes}.
These tools, typically based on static analyses, impose an additional complexity to
application development that is often non-trivial.
In our approach, the programmer specifies the degree of concurrency allowed and which
database constraints should be maintained -- the system enforces the specified
concurrency while trying to minimize coordination.
Some systems, such as Oracle multi-master replication, allow programmers to 
specify how to handle conflicting updates. Our approach is more complete, by
addressing a wider range of database constraints, which are key for enforcing
application correctness.

Many authors have proposed to relax applications consistency and tolerate temporary inconsistencies
in order to provide good performance at a planetary-scale~\cite{sagas,cap12years,quicksand}.
We follow the same principle, however we only allow programmers to specify concurrent semantics
when operations can be merged without affecting the integrity of the database.
To implement the proposed concurrent semantics, we use CRDT~\cite{crdts-sss} data types. These
data types allow merging concurrent operations without loss of updates, which are key
to implement some of the conflict resolutions that we propose.

AntidoteDB~\cite{antidote} is the backing store for AQL. AntidoteDB provides a key-object interface with
support for CRDTs and escrow data types that we used to implement the SQL semantics.  
AntidoteDB ensures Parallel Snapshot Isolation. A number of systems provide equivalent semantics~\cite{walter}.
AQL parallel-snapshot isolation \cite{walter}
with integrity invariants, in a similar way as snapshot isolation has been extended 
with integrity invariants \cite{Lin09Snapshot}.
Our approach for enforcing referential integrity cab be seen as a runtime version of
our previous work, IPA \cite{Balegas18IPA}, where, following a static analysis process, application
operations were modified in a way that guarantees that invariants are preserved 
when executed under weak consistency. In this work, we apply a similar idea in
runtime to SQL code.
Our approach can also be seen
as an extension of the approach to enforce serializability under snapshot isolation
proposed by Cahill et. al. \cite{Cahill:2008:SIS:1376616.1376690}, be executing
additional updates to force concurrency detection, and using conflict resolution
policies to achieve the intended behavior.

\section{Conclusion}
\label{sec:conclusion}

Programmers enjoy SQL's expressive data description and data access capabilities and
consistency model.
With \namebig{}, we provide a way to allow programmers to specify when and 
how to relax SQL consistency, while keeping the declarative data model and enforcing 
database constraints, including primary, check and foreign key constraints.
\namebig{} emphasizes the need for a well-structured database scheme that includes
database constraints with \name{} yielding customizable concurrency semantics.
With \namebig{}, we expect relaxing consistency to become less complex when compared to 
other highly available, geo-replicated key-value stores, resulting in safer programs.

Antidote SQL is open-source \cite{aql:code} and, besides the mechanisms described in this
paper, includes an indexing mechanism for primary and secondary keys. 
The values of the index are kept consistent with the data in the database, even
in the presence of concurrent updates that are solved using the conflict-resolution
policies defined for each table and table columns.
The preliminary evaluation of the system \cite{aql:msc} shows that the overhead of the mechanism to 
enforce foreign keys using an optimistic approach is negligible for insert and
update operations, but not for delete operations. For the \emph{delete-wins} policy,
there is also overhead related with the execution of select operations.

\section*{Acknowledgments}
We thank the anonymous reviewers for their comments that helped improving the
paper.
This work was partially supported by EU H2020 LightKone project (732505),
%%    \href{http://syncfree.lip6.fr/}{609\,551 SyncFree} (2013--2016), %
%    %\usebox{\EUflagBW},
and FCT/MCTES grants SFRH/ BD/87540/2012, UID/CEC/04516/2013, UID/CEC/50021/2013, 
Lis\-boa-01-0145-FEDER-032662 /PTDC/CCI-INF/32662/2017, and PTDC/CCI-INF/32081/2017.
%Computing resources were provided by an Amazon
%Web Services (AWS) in Education Research Grant.

%Programmers enjoy SQL's expressive data description and data access capabilities.
%With \namebig{}, we provide a concise declarative SQL dialect with semantics adapted to PSI consistency that allows to enforce typical constraints, including primary and foreign key constraints, in systems with high availability.
%To this end, we extended the DDL with annotations for automatic conflict resolution strategies on rows and individual cells.
%\namebig{} emphasizes the need for a well-structured database scheme with AQL yielding customizable concurrency semantics.
%With \namebig{}, we expect weak consistency semantics to become less complex to handle when compared to other highly available, geo-replicated key-value stores, resulting in safer programs.
%
%We are currently finalizing a prototype implementation of \namebig{} based on multi-level locks.
%Further work in progress is on the support for efficient queries and secondary indexes.

%\section*{Acknowledgments}
%Removed for anonymity.

%This research is supported in part
%    by EU FP7 SyncFree project (609551),
%%    \href{http://syncfree.lip6.fr/}{609\,551 SyncFree} (2013--2016), %
%    %\usebox{\EUflagBW},
%    FCT/MCT SFRH/BD/87540/2012, PTDC/ EEI-SCR/ 1837/ 2012 and PEst-OE/ EEI/ UI0527/ 2014. The research of Rodrigo\ Rodrigues is supported by the European Research Council under an ERC Starting Grant.

%\bibliographystyle{abbrv}
%\bibliography{bib}

\end{document}